\documentclass[acmsmall,author]{acmart}

 \usepackage{float}
 \usepackage{graphicx}
\usepackage{wrapfig}

\usepackage{amssymb}
\usepackage{booktabs}
\usepackage{xcolor}
\usepackage[normalem]{ulem}
\usepackage[table,xcdraw]{xcolor}
\usepackage[most]{tcolorbox}
\usepackage{enumitem}
\usepackage[final,commandnameprefix=always]{changes}

\tcbuselibrary{listingsutf8}

\usepackage{etoolbox}

\usepackage[normalem]{ulem} %
\usepackage{ifthen}

\newboolean{showrevisions}
\setboolean{showrevisions}{false}  %

\newboolean{showrevisions3rd}
\setboolean{showrevisions3rd}{false}  %

\definecolor{revadd}{RGB}{0,200,64}   %
\definecolor{revdel}{RGB}{200,0,0}    %
\definecolor{revnote}{RGB}{0,128,0}   %

\newcommand{\rev}[1]{%
  \ifthenelse{\boolean{showrevisions}}%
    {\begingroup\color{revadd}#1\endgroup}%
    {#1}%
}
\newcommand{\thirdrev}[1]{%
  \ifthenelse{\boolean{showrevisions3rd}}%
    {\begingroup\color{revadd}#1\endgroup}%
    {#1}%
}

\newcommand{\revdel}[1]{%
  \ifthenelse{\boolean{showrevisions}}%
    {\begingroup\color{revdel}\endgroup}%
    {}%
}

\newcommand{\revnote}[1]{%
  \ifthenelse{\boolean{showrevisions}}%
    {\marginpar{\footnotesize\color{revnote}{#1}}}%
    {}%
}

  {\ifthenelse{\boolean{showrevisions}}{\color{revadd}}{}}
  {}

\newtoggle{comments}
\toggletrue{comments}

\iftoggle{comments} {
\newcommand{\warren}[1]{{\color{blue}{WP: #1}\normalfont}}
\newcommand{\fanny}[1]{{\color{purple}{FC:#1}\normalfont}}
\newcommand{\tony}[1]{{\color{cyan}{TT:#1}\normalfont}}
\newcommand{\JensEmil}[1]{{\color{magenta}{JG:#1}\normalfont}}

}{
  \newcommand {\warren}[1]{}
  \newcommand {\fanny}[1]{}
  \newcommand{\tony}[1]{}  
  \newcommand{\JensEmil}[1]{}  
 }

\newcommand{\pid}[1]{%
  \ifcase#1 %
    \relax %
  \or %
    P9%
  \or %
  \or %
    P1%
  \or %
  \or %
    P2%
  \or %
    P3%
  \or %
    P4%
  \or%
    P5%
  \or%
    P10%
  \or%
    P11%
  \or%
    P6%
  \or %
    P12%
  \or %
    P7%
  \or %
    P13%
  \or %
  \or %
    P8%
  \else %
    Undefined %
  \fi
}

\newcommand{\bpstart}[1]{\vspace{1mm}\noindent\textbf{#1}}
\renewcommand{\paragraph}[1]{\bpstart{#1}}

\newtcolorbox{tcbsiderules}[1][]{blanker, breakable, 
     left=3mm, right=3mm, top=1mm, bottom=1mm,
     borderline vertical={1pt}{0pt}{black},
     before upper=\indent, parbox=false, #1}

\makeatletter
\newcommand{\reportquote}[2]{%
  \noindent
  \begin{tcolorbox}[
    blanker,    
    sharp corners,
    boxsep=0pt,
    left=10pt,
    right=6pt,
    top=0.2em,
    bottom=0pt,
    before skip=5pt,
    after skip=5pt,
    overlay={
      \draw[lightgray, line width=2pt]
        ([xshift=0.5em]frame.north west) -- ([xshift=0.5em]frame.south west);
    }
  ]
  \parbox{\dimexpr\linewidth-1em}{
    \small\sffamily #1 \mbox{\textsc{-- \textbf{#2}}}
  }
  \end{tcolorbox}
}
\makeatother

\makeatletter
\newcommand{\reportimplication}[1]{%
  \@ifundefined{color@imessagegray}{
    \definecolor{imessagegray}{RGB}{229,229,234}
  }{}%
  \noindent
  \begin{tcolorbox}[
    colback=imessagegray,
    colframe=gray!30,
    boxrule=0.2pt,
    arc=2mm,
    left=0.5em,
    right=0.5em,
    top=0.1em,
    bottom=0.1em,
    boxsep=0.3em,
    parskip=0pt,
    before skip=2pt,
    after skip=2pt,
    fontupper=\small\sffamily,
  ]
  #1 
  \end{tcolorbox}
}
\makeatother

\makeatletter
\newcommand{\reportassumption}[1]{%
  \@ifundefined{color@imessageblue}{
    \definecolor{imessageblue}{RGB}{174,198,207}
    \definecolor{imessagegray}{HTML}{F9F9F9}
  }{}%
  \noindent
  \begin{tcolorbox}[
    colback=imessagegray,
    colframe=gray!30,
    boxrule=0.2pt,
    arc=2mm,
    left=0.5em,
    right=0.5em,
    top=0.1em,
    bottom=0.1em,
    boxsep=0.3em,
    parskip=0pt,
    before skip=2pt,
    after skip=2pt,
    fontupper=\small\sffamily,
  ]
  #1 
  \end{tcolorbox}
}
\makeatother

\newcommand{\miniskip}{\par\vspace*{0.4em}\noindent}

\newcommand{\tinyskip}{\par\vspace*{0.2em}\noindent}

\AtBeginDocument{%
  \providecommand\BibTeX{{%
    \normalfont B\kern-0.5em{\scshape i\kern-0.25em b}\kern-0.8em\TeX}}}

\copyrightyear{2026}
\acmYear{2026}
\acmConference[CSCW '26]{CSCW 2026}{October 13--17, 2026}{Salt Lake City, UT, USA}
\acmBooktitle{Proceedings of the ACM on Human-Computer Interaction (PACM HCI) Journal}
\acmPrice{}

\acmDOI{XXXXXXX.XXXXXXX}

\acmISBN{978-1-4503-XXXX-X/18/06}

\begin{document}

\title[\emph{FaceValue}]{\emph{FaceValue}: Exploring Real-Time Self-View Overlays to Prompt Meaning-Oriented Self-Awareness in Remote Meetings
}

\author{Gun Woo (Warren) Park}
\email{warren@dgp.toronto.edu}
\affiliation{%
\institution{Department of Computer Science, University of Toronto}
\streetaddress{40 St. George St.}
\city{Toronto}
\state{Ontario}
\country{Canada}
}

\author{Anthony Tang}
\email{tonyt@smu.edu.sg}
\affiliation{%
\institution{School of Computing and Information Systems, \\Singapore Management University}
\city{Singapore}
\country{Singapore}}

\author{Fanny Chevalier}
\email{fanny@cs.toronto.edu}
\affiliation{%
\institution{Department of Computer Science, University of Toronto}
\city{Toronto}
\country{Canada}
}

\renewcommand{\shortauthors}{Park, Tang, and Chevalier}

\begin{abstract}

In remote video meetings, visual non-verbal cues, such as facial expressions or head movements, are seen continuously but often only partially. This increases ambiguity compared to in-person settings and can cause misinterpretation or misalignment between intended and perceived meaning. Motivated by communication theories, we designed \emph{FaceValue}, a technology probe that augments the self-view with private, real-time overlays. These overlays are subtle, suggestive prompts intended to help attendees reflect on how their cues might be interpreted by others. To invite personal interpretation, \emph{FaceValue} avoids behavioral labeling and instead aims to support meaning-oriented self-awareness: recognizing when visible cues may unintentionally (mis)communicate intent. We deployed \emph{FaceValue} in the wild with thirteen knowledge workers over multiple weeks, capturing perceived changes in self-awareness and behavior, and impressions on the design concepts, as self-reported by participants through diary entries and exit interviews. Participants felt \emph{FaceValue}
increased their awareness of potentially misaligned cues and motivated in-meeting adjustments, which they believe resulted in improved communication with other attendees.
We contribute a conceptual framing that positions visual non-verbal cues as a manipulable communication resource, a technology probe that aims to foster meaning-oriented self-awareness, and empirically-grounded design insights for future meeting systems.
\end{abstract}



\maketitle
\section{Introduction}

When attending remote meetings, we continuously convey meaning not just through what we say, but also through how we look. Our facial expressions, head movements, and gestures are %
on display, and other attendees use these visual non-verbal cues, like a nod, a frown, or a smile, to interpret how we feel and think~\cite{barnlund1970transactional}, %
decide how to respond, and shape the %
conversation~\cite{mehrabian1967inference,mehrabian1967decoding,walther1992interpersonal}.

Remote meetings make it more difficult for attendees to interpret one another's visual non-verbal cues. Small video tiles and cropped views of the body can make some cues easy to miss or misread~\cite{daft1986organizational}. When attendees encounter incomplete or ambiguous cues, they fill in the gaps, sometimes incorrectly~\cite{heider1958,kelley1967attribution,berger1974some}. This can hinder collaboration: a frown might cause someone to prematurely dismiss a good idea, or a lack of visible reaction might falsely convey agreement. These ambiguities and misinterpretations can erode trust and communication~\cite{park2020beyond}. %
Given the prevalence of remote meetings~\cite{zoomminutes2024}, we ask: \textit{How can we assist meeting attendees in enhancing their self-awareness of their own non-verbal behaviors in remote meetings?}

The self-view, a built-in feature in most remote video meeting tools, gives  attendees the ability to see their own video feeds alongside those of others. Seeing oneself can support self-monitoring~\cite{duval1972self}; however, there are two aspects to consider. First, individuals often hold the mistaken belief that others can readily perceive their internal emotional states---a phenomenon known as the Illusion of Transparency~\cite{gilovich1998illusion}. Second, when attendees use the self-view, they tend to focus more on managing their appearance than on what their expressions are communicating~\cite{leong2021exploring, chen2021afraid}. Our work aims to help attendees notice and reflect on what their cues might be communicating to others, which we characterize as \emph{meaning-oriented self-awareness}. With better awareness, attendees can choose to adjust their cues or clarify verbally when needed.

Prior work has shown that systems that highlight certain visual non-verbal cues can help individuals become more aware of their own non-verbal behavior. Chow et al.~\cite{r6_chow2025} introduced a sidebar interface that notified users of variations to (and instances of) low-level non-verbal behaviors like smiling, nodding, gaze, and posture. This real-time information helped users become more aware of their self-presentation in remote meetings. 
While Chow et al. focused on monitoring low-level behaviors from the user’s own perspective, we take a complementary approach by asking a self-reflective question: how might others perceive these behaviors, particularly in terms of communicative intent? Our work extends this trajectory by not only detecting the presence of cues but also surfacing their likely interpretations. This real-time feedback helps users notice when they may be sending unintended signals and adjust their behavior in the moment.
Our goal is to support people in maintaining this \emph{meaning-oriented self-awareness}.

To explore the concept of meaning-oriented self-awareness, we propose to augment the self-view with private, real-time overlay elements that highlight visual non-verbal cues when those cues may communicate meanings to others. To realize this approach, we derive design goals (Section \ref{researchFramework}) grounded in communication and design theories (Section \ref{sect:background}): detecting non-verbal cues that might convey specific meanings, and through real-time overlays, encouraging users to reflect on these~behaviors. 

To examine how meaning‑oriented self‑monitoring influences attendees' awareness and adjustment of non‑verbal cues across varied meeting contexts, we implement these design goals through a technology probe, \emph{FaceValue}, that analyzes facial expressions and head movements, and augments the self-view with private comic-style overlays that can evoke attendees' own meaning-oriented self-awareness (Section \ref{sec:implementation}). We deployed \emph{FaceValue} in an in‑the‑wild study with 13 knowledge workers 
(Section \ref{sec:experiment}). Participants used \emph{FaceValue} as part of their regular remote meetings, professional or casual. 
Our focus is on the perceived impact of \emph{FaceValue} on one's awareness and behavior as self-reported by participants, collected through self-administered diary entries and post-study semi-structured interviews.
Participants used \emph{FaceValue} for five meetings, over the course of one to three weeks. Our study focused on the following questions:

\begin{enumerate}[leftmargin=0.4cm, rightmargin=0.4cm]
    \item[] \textbf{RQ1:} How does real-time feedback on the potential interpretations of one’s visual non-verbal cues help remote meeting attendees \emph{notice} when their behavior may align or misalign with their communicative intent?
    \item[] \textbf{RQ2:} How do attendees describe \emph{acting} on meaning-oriented feedback in order to realign their visible cues with their communicative intent?
    \item[] \textbf{RQ3:} How do meeting characteristics (e.g., stakes, formality) shape attendees’ perceptions of the feedback’s usefulness and potential for distraction?
\end{enumerate}

Participants felt that using \emph{FaceValue} helped them observe their self‑view more intentionally. They %
reported feeling more able to notice moments when their on‑screen behavior might misalign with their intended messages. 
They also emphasized that \emph{FaceValue} was most useful during high-stakes or goal-oriented meetings, where the perceived importance of communicative clarity was higher. 
These findings suggest that private, metaphorical overlays can give attendees a provisional reference point for monitoring and adjusting their visual non-verbal behavior, supporting meaning-oriented self-awareness without imposing prescriptive feedback (Section \ref{sec:findings}). 

Our contributions are:

\begin{enumerate}
\item Conceptual framing. We frame visual nonverbal cues in video as a way to reinforce intended meaning, while recognizing that they may also unintentionally convey unintended meaning. Through this conceptualization, we motivate the need for meaning-oriented self-feedback.
\item \emph{FaceValue}, a lightweight technology probe that aims to foster meaning-oriented self-awareness and reflection through private, comic-style overlays.
\item Empirical insights from an in-the-wild deployment with 13 participants, indicating participants feel that meaning-oriented feedback can help them align their expressions with their intentions. We discuss scenarios where FaceValue is most effective, and based on a reflection about potential unintended consequences, we outline implications for designing future context-sensitive meeting tools.
\end{enumerate}
 \section{Background}
\label{sect:background}
We first outline theories of communication that ground our work  (\autoref{tab:theory-overview}). Then, we discuss relevant tools and current practices for self-monitoring in remote meeting, with a focus on the self-view. We also review prior systems supporting interpretation of visual non-verbal cues.

\subsection{Theoretical Grounding: Visual Non-verbal Cues in Remote Meetings}
\label{sec:theories}
Visual non-verbal cues such as facial expressions, gestures, and head movements are integral part of communication~\cite{barnlund1970transactional} (\autoref{tab:theory-overview}--T1). These cues indicate understanding, enrich verbal explanations, and convey attitudes toward the conversation content~\cite{10.1145/166266.166289,adler2006understanding}. To support effective communication, visual non-verbal cues should ideally be both clear and congruent with one's intended message. However, such cues are inherently ambiguous (\autoref{tab:theory-overview}--T5,T6): different observers can derive different meanings from the same cue~\cite{adler2006understanding,aviezer2017inherently}, and these perceived meanings may not always align with one's communicative intent~\cite{waxer1977nonverbal,gonzalez2025trusting}. Because people tend to over-estimate how well their cues are understood by others (\autoref{tab:theory-overview}--T4), they may not consciously refine their cues. This can be a problem, as research suggests that individuals often rely on visual non-verbal cues to interpret communicative meaning, sometimes placing greater trust in them than in speech, especially when the two conflict~\cite{mehrabian1967decoding,mehrabian1967inference}.

\begin{table}[t]
 \caption{Overview of the theories that inform our perspective on meaning-oriented self-awareness. Visual non-verbal cues have significant roles in communication, yet remote meeting environments often degrade these cues. This uncertainty affects the \emph{perceiving} attendee, who may misinterpret ambiguous cues, and the \emph{cueing} attendee, who may be unaware of how their cues are being perceived. Our work identifies gaps supporting \textbf{cueing attendees'} self-awareness and draws on  theories and design lenses to guide our approach.}
    \label{tab:theory-overview}
    \footnotesize
    \centering
    \resizebox{\textwidth}{!}{
    \begin{tabular}{p{0.2cm}p{4cm}p{2.3cm}p{8.5cm}}
    \toprule \textbf{ID} &
    \textbf{Theory} & \textbf{Perspective} & \textbf{Role in this Work} \\
    \midrule
    T1 & Transactional Model of Communication ~\cite{barnlund1970transactional} & General Communication  & Explains why visual non-verbal cues remain relevant even when verbal cues are available. \\
    \midrule
    T2 & Media Richness Theory~\cite{daft1986organizational} & Meeting Medium &  Highlights that remote meeting media can hinder the effective conveyance of visual non-verbal cues, thereby reducing communication richness and potentially increasing ambiguity in the interpretation of those cues. \\
    \midrule
    T3 & Objective Self-Awareness ~\cite{duval1972self,wicklund1975objective}& \textbf{Cueing Attendee} & Describes how individuals focus on themselves during communication, affecting their awareness and adjustment of expressive behavior. \\
    \midrule
    T4 & Illusion of Transparency~\cite{gilovich1998illusion} & \textbf{Cueing Attendee} &  Explains how individuals overestimate how accurately others perceive their internal states and intentions. \\
    \midrule
    T5 & Social Information Processing~\cite{walther1992interpersonal} & Perceiving Attendee  & Describes how people adaptively extract and interpret cues from available information sources in mediated interactions. \\
    \midrule
    T6 & Uncertainty Reduction Theory~\cite{berger1974some} & Perceiving Attendee & Explains how individuals seek to reduce uncertainty by interpreting available cues, sometimes leading to over-interpretation or bias. \\
    \midrule
    T7 & Affect-as-Interaction~\cite{boehner2005affect} & Design Lens & Encourages designing for emotional expression as socially situated and interpretable, rather than as a static, classifiable signal. \\
    \midrule
    T8 & Ambiguity-as-Design-Resource~\cite{gaver2003ambiguity} & Design Lens & Positions ambiguity not as a flaw, but as a productive feature that invites reflection and deeper interpretation in user experience. \\
    \bottomrule
    \end{tabular}
    }
\end{table}

This ambiguity is even more pronounced in remote meetings due to reduced media richness (\autoref{tab:theory-overview}–T2). Video tiles are often small, crowded, or cropped, causing many subtle visual cues to be lost or only partially conveyed~\cite{park2020beyond}.
Because people are fundamentally motivated to reduce uncertainties about others' emotions or intentions, they actively engage in drawing meaning from these limited or unclear visual non-verbal cues (\autoref{tab:theory-overview}--T6), intentionally or not, as long as video is enabled~\cite{walther1992interpersonal}. The problem is that the gap-filling process associated with interpreting missing, partial, or leaked cues is often biased~\cite{heider1958,kelley1967attribution}, resulting in inaccurate conclusions and misjudgment.
Although misinterpretations are often overlooked and inconsequential, empirical studies demonstrate tangible harm. For example, a study showed that remote attendees exhibit more negative affect and a tendency to perceive hostile intents from others~\cite{park2020beyond}. Misattributions have also been shown to undermine team success~\cite{whiting2019did}. %

Efforts to reduce cue ambiguity in remote meetings include refined symbolic cues, such as emoji reactions, button-press feedback~\cite{aseniero2020meetcues}, or explicit turn-taking intent displays~\cite{hu2023openmic}. Another strategy is to stream only a single visual cue, such as gaze~\cite{r3_he2021gazechat}, or machine-interpreted emotional states~\cite{hautasaari2024emoscribe}. While both approaches reduce communicative noise, they also reduce richness. This tradeoff can be acceptable for simple communication tasks, but in more complex tasks, it can increase uncertainties ~\cite{daft1986organizational,berger1974some}. Professionals' preference for videos~\cite{kornferryCamerasOff}---hence visual non-verbal cues---underscores the importance of supporting this form of communication in remote meetings.

Our work takes a different approach to reducing cue ambiguity. Motivated by Chow et al.'s suggestion that ``the evaluation of one's own non-verbal behavior from the perspective of the other attendees [could] inform actionability''~\cite{r6_chow2025}, we provide private feedback that shows how a participant's cues \emph{might} be interpreted by others. We emphasize that the feedback reflects \emph{possible} meanings inferred from automated analysis of the attendee's own cues, and does not incorporate actual data from other attendees' interpretations during the meeting, as these remain internal. This feedback is designed to help individuals consider potential (mis)alignments with communicative intent, leaving them free to decide whether any adjustment is needed.  %
In other words, our objective is to encourage self-assessment and deliberate action toward reducing ambiguity at the source by evoking meaning-oriented self-awareness in the cueing attendee (\autoref{tab:theory-overview}–T3).

\subsection{Self-Monitoring in Remote Meetings}
When interacting with others, we naturally engage in \emph{self-monitoring}, i.e. aligning our expressive behavior with social expectations and communicative goals~\cite{snyder1974self}. In this paper, we use the ``self-monitoring'' construct in a situated, behavioral sense, i.e., moment-to-moment attention to and adjustment of one's expressive cues during interaction~\cite{leshed2010visualizing,meyer2017design}. This is related to, but distinct from the original psychological construct of self-monitoring as a stable \emph{trait}~\cite{snyder1974self}, which describes individual differences in how people monitor their social context, and adapt their self-presentation, but does not refer to any specific behavior (e.g., literally looking at one's own face more often). We return to this distinction in \autoref{sect:futurework} when considering how trait self-monitoring may shape people's experiences with our system. In this behavioral framing, self-monitoring goes hand in hand with \emph{self-awareness} of expressive cues. Consistent with this view, some studies suggest that awareness of one's expressions can help calibrate perceived demeanor~\cite{duval1972self} and user control~\cite{r1_seitz2024mirror}, and thus enhance clarity, rapport~\cite{sypher1983perceptions,verheijden2023developing}, and relational communication~\cite{r2_miller2017through}.

In remote meetings, the self-view panel, focused on the face of attendees, can facilitate this process, by enhancing self-awareness and allowing real-time adjustment of visual non-verbal cues. However, prolonged exposure to one's own image through the self-view can lead to fatigue, anxiety~\cite{bailenson2021nonverbal,shockley2021fatiguing,silence2022our}, discomfort~\cite{de2009image}, or cognitive load~\cite{r1_seitz2024mirror}. 
These effects vary across individuals~\cite{kuhn2022constant} and may not arise at all for some~\cite{li2023filters,k2021meeting}. For others, however, aspects of the self-view can undermine its benefits or introduce additional challenges.

Psychology often frames self-awareness as a component of mindfulness, typically associated with well-being~\cite{sutton2016measuring,vago2012self,gu2015mindfulness}, which contrasts with reports that self-view may induce stress~\cite{riedl2022stress}. There are reports that self-views often result in monitoring for more cosmetic cues e.g., hair, background~\cite{chen2021afraid,leong2021exploring}, which serves little function in communication.
Building on these insights, we hypothesize that focusing on communicative cues (e.g., expressions of agreement) may support more purposeful interaction with the self-view. Our work explores this approach by augmenting the self-view with feedback on potentially communicative visual non-verbal cues (e.g., subtle displays of agreement or confusion), as a mechanism to supporting attendees in recognizing, interpreting, and adjusting their own cues in remote meetings.

While not our primary focus, we note that leveraging the self-view also allows to preserve other benefits this view affords, such as coordinating objects relevant to conversations within the camera's frame~\cite{r4_hu2023thingshare, liu2023experiencing,liao2022realitytalk,r7_hall2022augmented,park2024jollygesture}. This aspect of communication is important and complementary to our focus on visual non-verbal cues, albeit beyond our scope.

\subsection{Visualizing Visual Non-verbal Cues}

Many remote collaboration tools aim to visualize participants' emotional or engagement states to help interpret group dynamics. These include emoji reactions, active-speaker highlighting~\cite{murali2021affectivespotlight}, and experimental dashboards~\cite{aseniero2020meetcues,das2022cannot,zeng2020emotioncues,lee2024investigating,maeda2022calmresponses,ez2020emodash,benke2022teamspiritous}. While helpful for group awareness, such interfaces offer limited support for \emph{personal} self-monitoring. Feedback is often public, encouraging shared interpretation rather than private reflection. Our work introduces a personal reflective tool to support individual self-monitoring.

When dealing with visual non-verbal cues, there are two design approaches that we can consider. Under the affect-as-information paradigm~\cite{boehner2005affect}, visual non-verbal cues are treated as classifiable data~\cite{r6_chow2025,rojas2022towards,aseniero2020meetcues,das2022cannot,zeng2020emotioncues,lee2024investigating,ez2020emodash,benke2022teamspiritous}. Visualizations here are often literal, using labels or percentage scores presenting discrete, context-free observations, such as ``three head nods in the past two minutes’’ or ``a 55\% probability of smiling.’’ Such systems often still avoid prescriptive \emph{guidance} (e.g., ``nod twice now’’), and instead adopt non-normative stances for feedback, which leaves the action plan to the users. For example, Chow et al.~\cite{r6_chow2025} presents counts of detected behaviors like smiles or nods, but leaves interpretation entirely to the users, to preserve authenticity in expressions. This paradigm supports analytical perception of facial expressions~\cite{chen2010contribution,bradshaw1971models,konar2010holistic,sekuler2004inversion}, but risks over-confidence in Facial Expression Recognition (FER) accuracy~\cite{rojas2022towards,zhai2024effects}.

By contrast, the affect-as-interaction paradigm (\autoref{tab:theory-overview}--T7) treats visual cues as meaningful through interaction. For instance, AffectiveSpotlight~\cite{murali2021affectivespotlight} enlarges video feeds rather than labeling them, and other systems~\cite{howell2016biosignals,rajcic2020mirror} evoke reflection through ambient feedback. This paradigm is not inherently superior to affect-as-information; each supports different user needs. If the goal is quick, quantifiable awareness, the information paradigm may work better. However, if the aim is to support users in reflecting on how their cues might be perceived holistically and contextually, the interaction paradigm offers a more interpretive lens~\cite{boehner2005affect}. We adopt this latter approach to evoke user’s holistic awareness on not just what cues occurred, but how those cues may align, or misalign, with intended meaning. Facial expressions are often perceived holistically~\cite{calder2005configural,calder2000configural,tanaka2012mixed}, so prompting reflection on multiple integrated cues can be important. Designing for ambiguity, via color gradients, abstract visuals, or symbolic shapes~\cite{csemsiouglu2022emote,marino2024cohere}, can support this paradigm by encouraging interpretation. We similarly use ambiguity as a resource for design (\autoref{tab:theory-overview}--T8) to stimulate attendees' reflection.

Closest to our work is the \emph{Novecs} system that automatically detects and displays users' visual non-verbal cues, to support workers' self-presentation in remote meetings~\cite{r6_chow2025}. Our work shares similar design choices: providing real-time personal feedback in a non-normative way, which Chow et al.'s experiment showed can enhance perceived self-awareness of the participants~\cite{r6_chow2025}. Motivated by these findings, we extend this line of research by exploring two novel aspects, drawing on theories of communication pertaining to ambiguity in cues (Section \ref{sec:theories}). First, \emph{Novecs} displays feedback in glanceable sidebar of facial and postural cues---a form of separate dashboards~\cite{lee2024investigating,zeng2020emotioncues}. Here, we explore in-situ overlays, as they may help attendees connect feedback with real-time cues~\cite{pohl2024body}. Second, \emph{Novecs} focused on smiling, nodding, gaze, and posture in isolation. Our work looks into ways to enable more holistic interpretation of one's own visual non-verbal cues, with respect to how others may (mis)interpret the meaning behind these cues---a future direction suggested by Chow et al.~\cite{r6_chow2025}. Finally, Chow et al.'s longitudinal study focused on work meetings. We follow a similar methodology, but invited participants to experiment with our probe in meetings of their choosing, allowing us to expand insight on the type of meetings where personal meaning-oriented feedback might be valuable.

\subsection{Recognizing Visual Non-verbal Cues}
To provide feedback, we need to detect visual non-verbal cues and infer their potential meanings. Prior work in remote communication has largely focused on verbal cues~\cite{hillard2003detection,wang2011identifying}, but visual non-verbal cues, especially facial expressions, are also key~\cite{barnlund1970transactional}. Advances in affective computing have yielded Facial Expression Recognition (FER) systems that can match or surpass untrained human judges on constrained benchmarks~\cite{li2020deep}. Seminal work comparing posed versus spontaneous expressions~\cite{motley1988facial} warns, however, that benchmark superiority may not transfer to natural settings. FER has been applied in education~\cite{zeng2020emotioncues,sun2018exploration,10.1145/3383923.3383949}, mental health~\cite{veerakannan2024affective,jiang2022utilizing}, and robotics~\cite{liu2017facial,khan2024human}.

FER commonly uses categorical (e.g., happiness, anger~\cite{ekman1992facial}) or dimensional (valence–arousal~\cite{russell1980circumplex}) models. Categorical models suit users seeking direct feedback~\cite{csemsiouglu2022emote}, while dimensional models capture nuance. However, FER accuracy depends on training data, which may lack real-world complexity~\cite{mollahosseini2017affectnet,kanade2000comprehensive,livingstone2018ryerson,park2020differences}, and systems still lack contextual understanding. Privacy, demographic-bias, and other ethical concerns persist~\cite{asherschapiro2022Zoom,rhue2018racial}, prompting some companies to discontinue FER services (e.g., Microsoft's 2022 withdrawal of emotion recognition from Azure Face~\cite{microsoftMicrosoftsFramework}).

To balance ambiguous, reflective feedback~\cite{gaver2003ambiguity,boehner2005affect} with these challenges, we display FER output only to the attendee, not to others. This private feedback helps attendees refine expression with reduced risk of misinterpretation. We incorporate categorical and dimensional outputs to support varied preferences. We focus on facial expressions and head movements due to their salience in communication~\cite{adler2006understanding,seidela2010impact, helweg2004nod}. Future work may expand to hand movements or posture.
\section{Design Goals}
\label{researchFramework}

Drawing on theory, we identify two insights that motivate our approach and inform the design goals that follow. (1) People tend to overestimate how clearly their internal states are understood (\autoref{tab:theory-overview}-T4). We hypothesize that presenting feedback supporting this understanding could help, since theory suggests that (2) making oneself an object of attention supports regulative self-monitoring (\autoref{tab:theory-overview}-T3). To achieve this, we reconceptualize the self‐view as an active, reflective communication aid that supports meaning-oriented self-awareness. We articulate concrete design goals:

\begin{itemize}
  \item \textbf{DG1: Detect and Surface Visual Non-verbal Cues that Convey Meaning.} Actively detect and highlight visual non-verbal cues most salient to conversational meaning, to help users gauge how others may interpret their non‐verbal cues. The purpose of doing this is to re-center users' attention on communicative aspects of their appearance, rather than cosmetic self-checks~\cite{chen2021afraid,leong2021exploring}. Our implementation focuses on facial expressions and head movements—such as nods, frowns, and smiles—as these are cues interlocutors use to infer emotional state and stance~\cite{adler2006understanding,seidela2010impact, helweg2004nod}. 

  \item \textbf{DG2: Provide Subtle, Suggestive Feedback.}
  Invite user's interpretation and reflection without overstepping into definitive judgments about their internal state. Doing so guards against prescribing normative behavior~\cite{r6_chow2025}. Given that FERs are imperfect, %
  recognition results can vary.
  Thus, we treat its output as one plausible but fallible reading of visual non-verbal cues, rather than as
  ground truth
  . Informed by the affect‐as‐interaction paradigm~\cite{boehner2005affect}, we choose against rigid categorizations of emotional states and stance (e.g., ``you look angry'') that can reinforce overly-analytic self-monitoring. Instead, we realize this design following principles of ambiguous design~\cite{gaver2003ambiguity}, through suggestive display of the detected visual non-verbal cues that invites a wider range of interpretations. For instance, we could overlay gently suggestive visual hints for valence shifts within the self‐view. 

  \item \textbf{DG3: Sustain Real‐Time, In-Context, Continuous Feedback.}  
    Deliver immediate, ongoing, in-situ feedback about how visual non-verbal cues could be interpreted by others. Doing so enables in-the-moment evaluation, thereby making it possible for the cueing attendee to iteratively adjust their visual non‐verbal cues or offer verbal clarification in the flow of conversation, aligning more closely with their communicative intentions. We use in-situ self-view overlays to facilitate connections between feedback and one's real-time cues~\cite{pohl2024body}.
\end{itemize}

Applying these design goals, we turn the self-view into an \emph{active} mirror that draws attention to visual non-verbal cues when attention is needed (\textbf{DG1}). We provide suggestive overlays rather than prescriptive judgments on what these cues could mean to others (\textbf{DG2}), in-situ and in real-time (\textbf{DG3}). This meaning-oriented feedback approach aims to invite reflection, which could encourage attendees to iteratively adjust their visual non‐verbal cues to match their communicative intentions.%

\section{Design and implementation of the \emph{FaceValue} probe}
\label{sec:implementation}

\begin{figure}[t]
    \centering
    \includegraphics[width=\linewidth]{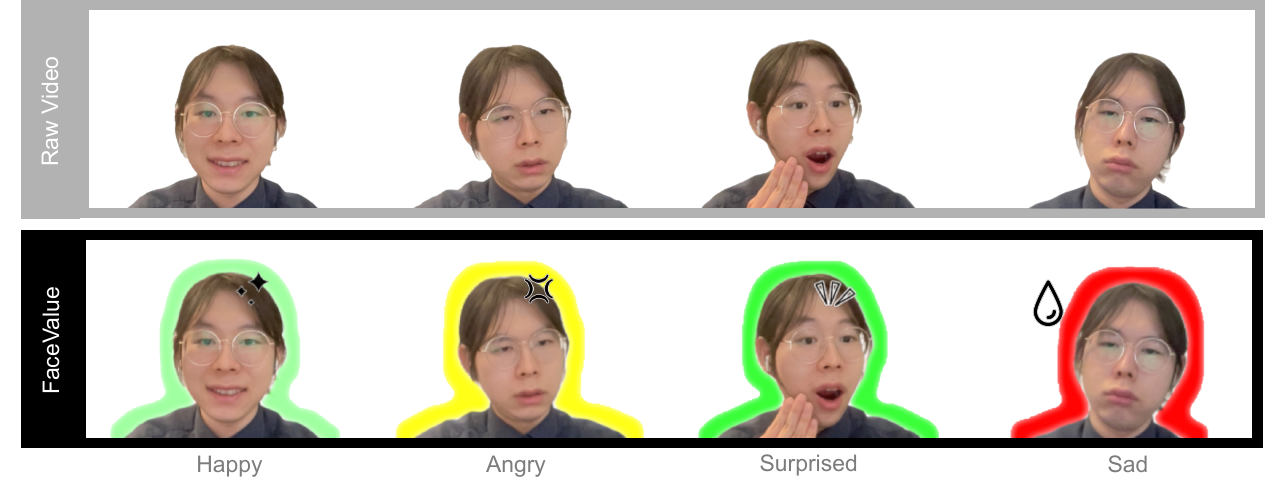}
    \caption{\emph{FaceValue} augments the user's self-view based on its interpretation of their communicative expressions. For example, when the user appears to be expressing agreement—expressed with a smile—their face is outlined in green. Conversely, when they appear to express disagreement—expressed with a frown—their face is outlined in yellow. Some specific expressions or head movements are accompanied by an icon. For instance, a smile is paired with a sparkle icon, while an angry face is paired with a popping vein icon. The possible interpretations, such as ``Happy'' or ``Angry'' are not visible to users.}
    \Description{Top row has raw images of happy, angry, sad and surprised human faces. Bottom row has corresponding processed facial images, as per the description on the caption, that is, happy face is emphasized by a green line, with a sparkle icon, angry face by a yellow line with a popping vein icon, surprised face by a green line and an exclamation icon, and sad face by a red line and a droplet icon.}
    \label{fig:emotion}
\end{figure}

We implemented \emph{FaceValue} as a proof-of-concept technology probe~\cite{hutchinson2003technology} to explore how people experience and incorporate meaning-oriented feedback%
. \emph{FaceValue} operationalizes our design goals (\textbf{DG1, DG2, DG3}) to explore meaning-oriented self-awareness. The system interprets users' communicative visual non-verbal cues (e.g. agreement, neutrality, confusion, and so forth) by detecting facial expressions and head movements, and augments the attendee's self-view so they are aware of how their expressions may be interpreted by others. For example, if a user smiles, a thick green band encircles their head in the self-view (\autoref{fig:emotion}), suggesting (but not prescribing) agreement or positivity. For certain expressions or head movements, the user's self-view will show overlaid icons or lines. For example, if a user is nodding, they will see speed lines at the top and bottom of the face (\autoref{fig:cues}). 
The system also monitors for sudden shifts in expression valence and alerts the user when a major change occurs---for example, when a smile or nod quickly turns into a frown or head shake.
Since these augmentations are only visible to the user in their self-view, our aim is to promote self-awareness of the potential meaning of their communicative expressions.

\subsection{An Example Scenario}
\begin{figure}[t]
\centering
\includegraphics[width=\linewidth]{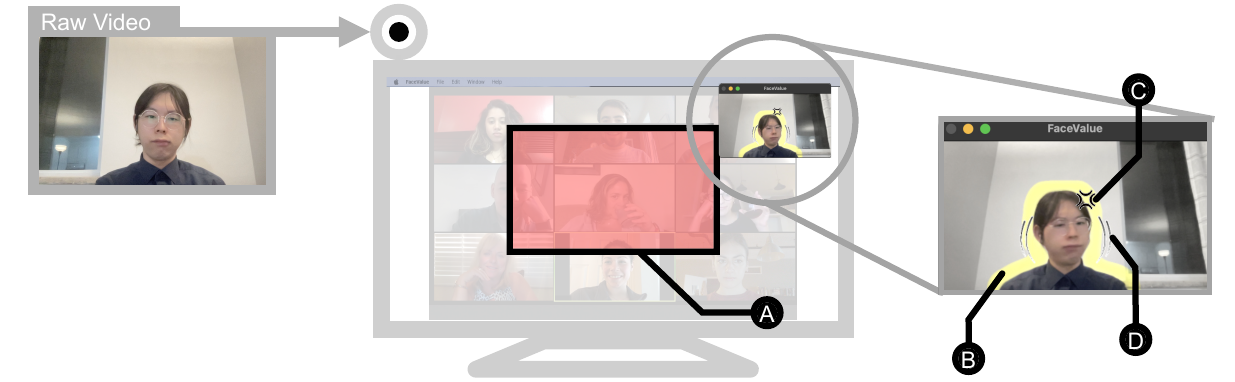}
\caption{Conceptual illustration of the augmented self-view in \emph{FaceValue}. (A) A brief red overlay signals a significant change in valence; (B) Yellow lines around the face; (C) Subtle icons hinting at possible interpretive cues (e.g., anger, surprise); (D) line overlays indicating minor movements that might otherwise be overlooked.}
\Description{On the left, a raw video feed from the webcam is displayed. At the center, a monitor shows a Zoom meeting window with a red, semi-transparent overlay positioned in the middle (labeled A). In the top-right corner of this window, there is a small FaceValue view. In this view, a person is outlined with a yellow boundary (labeled B) and features lines on either side of their face (labeled D). Additionally, a popping vein icon (labeled C) is shown near the individual.}
\label{fig:vis}
\end{figure}

To illustrate how the \emph{FaceValue} probe works, consider a graduate student, Taylor, joining a team meeting with Ash. Ash is presenting a new UI design idea for their project.

\paragraph{Moment 1: Unintended Negative Expression.}
When Taylor becomes briefly confused, they inadvertently frown for a moment. \emph{FaceValue} detects a shift to negative valence and briefly displays a red, semi-transparent overlay (\autoref{fig:vis} (A)). Looking at the augmented self-view, Taylor notices \emph{yellow lines} around their face and a ``popping vein'' icon (\autoref{fig:vis} (B, C)), suggesting that their expression may be interpreted by others as frustration or negativity. Realizing this might send the wrong signal, Taylor consciously relaxes their face, preventing a potential misunderstanding.

\paragraph{Moment 2: Subtle Agreement Expression.}
Later, Ash describes one of the ideas. Taylor nods slightly in agreement. \emph{FaceValue} identifies a mild motion and displays faint lines around Taylor's face (akin to \autoref{fig:vis}\,(D) but for nods), hinting that the nod may be too subtle for others to notice. On realizing this, Taylor amplifies the nod, signaling clear support to Ash.

In this scenario, \emph{FaceValue}'s private, nuanced cues in Taylor's self-view help them to refine how they project agreement or uncertainty. Rather than prescribing a ``correct'' expression, the system merely alerts the user of shifts in facial expressions and head movements (\textbf{DG2}), giving them the freedom to interpret and adjust as deemed fit (\textbf{DG1}, \textbf{DG3}). 

\subsection{Sensing: Recognizing Facial Expressions and Head Movements}

To direct users' attention toward meaningful visual non-verbal cues (\textbf{DG1}), \emph{FaceValue} first identifies facial expressions and head movements. We employ \emph{EmoFAN}~\cite{toisoul2021estimation}, a model that outputs categorical expressions (e.g., happy, angry) and dimensional expressions (valence/arousal).
EmoFAN achieves high accuracy on ranges of common datasets, including $\approx$75\% on AffectNet~\cite{mollahosseini2017affectnet} (categorical)\footnote{This model has been tested for other major datasets as well, and shows excellent performance.} and an RMSE of 0.29/0.27 for valence/arousal regression, which we consider sufficient to indicate expressive fluctuations in real-world usage. These raw detection labels are not visible to users. To detect head movements, we extract the 3D transformation matrix from \emph{MediaPipe}~\cite{lugaresi2019mediapipe} (pitch, roll, yaw) which allows us to capture behaviors such as nodding, shaking, or tilting.

Our implementation runs at 10+ FPS on an Apple M1 chip with minimal latency,providing real-time responsiveness (\textbf{DG3}).

\subsection{Visualization Mapping}
\begin{table}[t]
    \caption{Details of the \emph{FaceValue} mapping between recognition and the visual overlays }
    \label{tab:mapping}
\footnotesize
    \begin{tabular}{lll}
    \hline
       \textbf{Measure}  & \textbf{Input} & \textbf{Mapping}\\
       \hline
      Valence (positive vs. negative)   & V>0,A>0 & Outline colour: Green\\
      Arousal (intensity)  & V>0, A<0 & Outline colour: Blue\\
         & V<0, A<0 & Outline colour: Red\\
         & V<0, A>0 & Outline colour: Yellow\\
         & Euclidean distance & Outline opacity (opacity $\propto$ distance)\\
         \hline
        Detected Facial Expression & Happy & Sparkle icon\\
         & Sad & Droplet icon \\
         & Surprised & Exclamation icon\\
         & Angry & Popping vein icon\\
         \hline
         Head Movement & Nodding & Lines on the top and bottom of the face\\
         & Shaking & Lines on the left and right of the face\\
         &Tilting & Lines coming from the face\\
         \hline
         
    \end{tabular}
\end{table}
\begin{figure}[t]
    \centering
    \includegraphics[width=\linewidth]{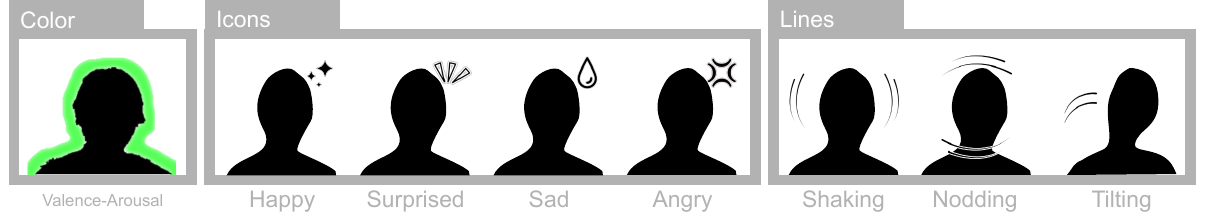}
    \caption{The ``Visualization Mapping'' phase produces three main forms of visual cues: color-based overlays (left), icons (center), and lines near the face (right).}
    \Description{Left image: green line around a face. Center image: icons such as sparkle, exclamation, droplet, and popping vein. Right image: lines placed around the face in various orientations.}
    \label{fig:cues}
\end{figure}

In the sensing stage, \emph{FaceValue} captures valence and arousal values, categorical emotion confidences, and head movement signals.
We map these values to visualization elements that aim to be subtle and suggestive (\textbf{DG2}) to encourage user-driven interpretation, as described in \autoref{tab:mapping}, and illustrated in \autoref{fig:cues}. Inspired by the visual language of comics/manga~\cite{forceville2017stylistics,shinohara2009pictorial}, our design lets users see hints of their expressions without receiving explicit emotion labels that can mislead.

\textbf{Valence \& Arousal (VA) $\Rightarrow$ Outline Color and Opacity.}
We adapt a hue scheme similar to prior work~\cite{kalateh2024towards}, where specific regions in VA space map to colors. For example, excited/happy tends toward green, sad/depressed toward red, angry/afraid toward yellow, and relaxed/calm toward blue. Intensity is encoded in the overlay's opacity, derived from the Euclidean distance from the VA origin (i.e., stronger expressions lead to higher opacity). To avoid rapid color flicker from frame-to-frame noise, we apply a Kalman filter~\cite{welch1995introduction} for smoothing.

\textbf{Detected Facial Expression $\Rightarrow$ Icons.}
When a detected emotion's confidence exceeds 80\%, \emph{FaceValue} displays an icon: sparkle (happy), droplet (sad), exclamation (surprise), or popping vein (anger); see \autoref{fig:cues} (center). While explicit, these icons are intentionally ambiguous rather than literal, consistent with \textbf{DG2}'s philosophy of inviting contextual interpretation. For example, a popping vein icon may suggest anger or frustration, but we omit textual labels; interpretation is left to users, aligning with Chow et al.'s non-normative feedback~design~\cite{r6_chow2025}.

\textbf{Head Movements $\Rightarrow$ Lines.}
Using pitch/roll/yaw data, we detect nods, shakes and tilts. For nods or shakes, we compute head position variance within a sliding window, accounting for their cyclic nature. If above a threshold, lines appear around the user's face (\autoref{fig:cues}, right). For tilts, we use the mean angle. Higher variance increases line opacity, signaling intensity. Thresholds can be tuned for different movement styles.

\subsection{Composition: Placing Visual Elements on the Attendees' Videos}

\emph{FaceValue} displays the icons, lines, and hue overlays directly over/near the user's face~\cite{pohl2024body} rather than off to the side~\cite{lee2024investigating,zeng2020emotioncues}. This design supports \textbf{DG2} by allowing immediate comparison between the raw facial expression and the abstract cue. Because the system avoids explicit emotion labels, the user retains interpretive flexibility (e.g., a popping vein icon might prompt them to think, ``Am I frustrated?'', ``Am I concentrating?'', or ``Did the model misinterpret my brow furrow?'').

A key design goal is to shift users from casual glances at their appearance to purposeful checks of communicative expressions. Conventional self-views do not explicitely support noticing significant changes in facial expression, so \emph{FaceValue} detects sudden valence shifts and briefly displays a semi-transparent overlay (\autoref{fig:vis}\,(A)). These short-lived ``flash'' signals prompt users to reflect: ``Am I aware my expression changed? Was it intentional?'' Additionally, to help the face stand out, we  reduce background saturation using MediaPipe~\cite{lugaresi2019mediapipe}, reinforcing focus on the face.

\section{Deployment Study}
\label{sec:experiment}

The goal of the study was to better understand how an augmented self-view might support users' self-awareness of their own communicative visual non-verbal cues during video conferencing. Specifically, we wanted to probe how people experience and interpret their own cues. To do so, we conducted an in-the-wild deployment study exploring how individuals integrated \emph{FaceValue} into real-world remote meetings, and how this shaped their perceptions, experiences, and any perceived changes in behavior or communication practices they felt emerged through~its~use. 

Following Chow et al.~\cite{r6_chow2025}, we opted for an in-the-wild deployment to address our research questions. Short-term laboratory experiments often oversimplify meeting dynamics, are prone to the novelty effects~\cite{elston2021novelty}, and are overly influenced by unique social dynamics~\cite{grudin1994groupware}, while lacking the time needed for users to integrate the tool into their routines~\cite{kjaerup2021longitudinal}. In contrast, in-the-wild studies excel at capturing real-life usage conditions, increasing ecological validity, and revealing genuine patterns of tool adoption and use~\cite{brown2011into}, which better align with our goals. %

Since our research focuses on self-awareness of the cueing attendee, we relied on internal measures, i.e. participants' self-reports, to probe the impact on their own experiences. We employed a diary study approach~\cite{houben2016physikit}, where participants, after having engaged in a meeting with \emph{FaceValue}, reported on their experiences using \emph{FaceValue} in a semi-structured way. We used this approach to capture participants' immediate reactions to the tool during deployment, and followed this up with a post-study interview to capture longer-term reflections. External measures, such as objective observations of behavior change, are challenging in this context because real-world meetings vary greatly in topics, attendees, and dynamics. Capturing meaningful patterns would require extensive controls and a consistent baseline of similar meeting contexts over time, which was beyond the scope of this study. We discuss the limitations of our approach in \autoref{sect:futurework}.

\subsection{Participants}
We recruited 15 participants via convenience sampling, snowballing, and advertisements on an undergraduate portal and a HCI Slack group. Two discontinued due to busy schedules, leaving 13 full participants. All but one used a Mac with Apple Silicon Chip\footnote{\pid{12} used an Intel iMac, which we confirmed provides equivalent experience. The minimum hardware was a MacBook Air with M1 Chip and 8GB of RAM.}, ensuring consistent performance for real-time video processing. All participants were fluent in English, 18 or older, and planned to attend at least five remote meetings over one to three weeks.

Participants included 5 women and 8 men, all Canada-based, with occupation spanning: business analyst (\pid{7}), assistant professor (\pid{3}), UX researcher (\pid{5}), postdoctoral researcher (\pid{11}), system adminstrator (\pid{12}), graduate (\pid{6}, \pid{8}, P8--11, \pid{14}) and undergraduate student (\pid{13}). Seven were aged\footnote{Collected in ranges.} 25--34 (P1--2, \pid{8}, P9--11, \pid{14}), four 18--24 (P3--4, P7--8), one 35--44 (\pid{11}), and one 45--54 (\pid{12}).

\subsection{\emph{FaceValue} Application}
\emph{FaceValue} was deployed as an executable macOS application running on participants' computers during their real meetings. All processing was done on-device with no logging. Participants used \emph{FaceValue} alongside their preferred software (e.g., Zoom, Microsoft Teams, FaceTime). The \emph{FaceValue} window was instructed to be placed over the standard self-view (moved and resized), allowing participants to receive real-time feedback on their facial expressions and head movements. 

\subsection{Procedure}
We conducted a three-phase diary study:
\begin{itemize}
\item \textbf{Phase 1 (Onboarding)}, $\sim$ 30min. In a Zoom session, participants provided informed consent, installed \emph{FaceValue}, were instructed on what \emph{FaceValue} does and how to use it, answered a pre-study questionnaire (demographics and general meeting experiences), and participated in a semi-structured interview about prior remote meeting experiences.
\item \textbf{Phase 2 (In-the-Wild Usage)}, 1-3 weeks. Participants used \emph{FaceValue} in five actual video conferencing meetings of their choice with respect to type (e.g. work related, casual meeting with friends, etc.), and size. After each meeting, they filled out a diary entry capturing key metadata (e.g., duration, topic, number of attendees) and describing any notable experiences they had while using \emph{FaceValue}; see section~\ref{sec:diary}. Not accounting for meeting time, we estimate participants spent 5min to fill out each diary.
\item \textbf{Phase 3 (Post-Study Interview)}, $\sim$ 30min. Participants completed a post-study questionnaire (including preference tests) and a semi-structured interview conducted on Zoom, about their overall experiences, focusing on perceived changes in communication and self-awareness of one's communicative visual non-verbal cues. During the interview, participants were asked to compare experience \emph{with} and \emph{without} \emph{FaceValue}.
\end{itemize}

Participants received a \$30 CAD gift card in compensation. The study was approved by our institutional ethics board and piloted to refine procedures before the main study. Study materials are provided in the supplemental.

\subsection{Diary instrument}
\label{sec:diary}
Participants were asked to complete a diary online after each meeting with \emph{FaceValue} via a personalized link with pre-filled participant ID. These diaries aimed to capture their immediate reflections on using \emph{FaceValue} in that meeting, and how their perceptions of \emph{FaceValue} evolved over time. \autoref{tab:diary} shows the questions in this instrument. 

\begin{table}[h]
 \caption{Questions in the diary, prompting short-form answers (with drop-down lists of possible values for Q1--6), Likert scores (strongly agree (1) to strongly disagree (5) for Q7--8), or long-form answers (Q9--13). }
    \label{tab:diary}
\footnotesize
    \centering
    \resizebox{\textwidth}{!}{
    \begin{tabular}{lll}
    \hline
       \textbf{Question}  \\
        \hline
      Q1: Date and time of the meeting  \\
      Q2: Type of the meeting \\
      Q3: Duration of the meeting  \\
       Q4: Usual frequency of the meeting  \\
       Q5: Number of attendees\\
       Q6: Perception of speaking more, less, or about the same during the meeting, compared to others \\
       Q7: Extent of agreement: I viewed myself more often \\
       Q8: Extent of agreement: I viewed others more often \\
       Q9: What were you trying to understand about yourself when you glanced at your \emph{FaceValue}? \\
       Q10: Were there moments when \emph{FaceValue}'s augmented self-view felt particularly helpful or distracting? \\
       Q11: In what ways could \emph{FaceValue} have helped (you) or changed this meeting? \\
       Q12: Did any attendee comment on your appearance, expressions, or demeanor in a way that might relate to your on-camera presentation? \\
        Q13: Any other thoughts or observations? \\
        \hline
    \end{tabular}
    }
        \vspace{-0.5em}
\end{table}

\subsection{Analysis}
\label{sec:methods-analysis}
We performed a thematic analysis of the semi-structured interviews transcripts and written diaries. 
We inductively captured themes relevant to our research questions, following an interpretivist approach to reflexive thematic analysis~\cite{braunThematicAnalysisPractical2022}. Our analysis focused on participants' explicit descriptions of their self-monitoring experiences and perceptions, considering presence, rather than prevalence, in line with this approach. First, the lead author reviewed the transcripts and diary entries in full, and performed open, line-by-line coding in MAXQDA\footnote{https://www.maxqda.com}. This step generated codes that captured salient themes and concepts in the participants' account. The lead author then iteratively reviewed and refined the initial codes, merging, splitting, and discarding codes as necessary with input from all authors in weekly discussions. The authors' cultural backgrounds span East Asia, Europe, and North America, and each has lived and worked in at least two countries across these regions (see Supplementary Material). This may have shaped our interpretive lens (e.g., meeting norms centered towards ``Western'' meeting styles) during analysis.%

\section{Findings}
\label{sec:findings}

\begin{figure}[h]
    \centering
    \includegraphics[width=\linewidth]{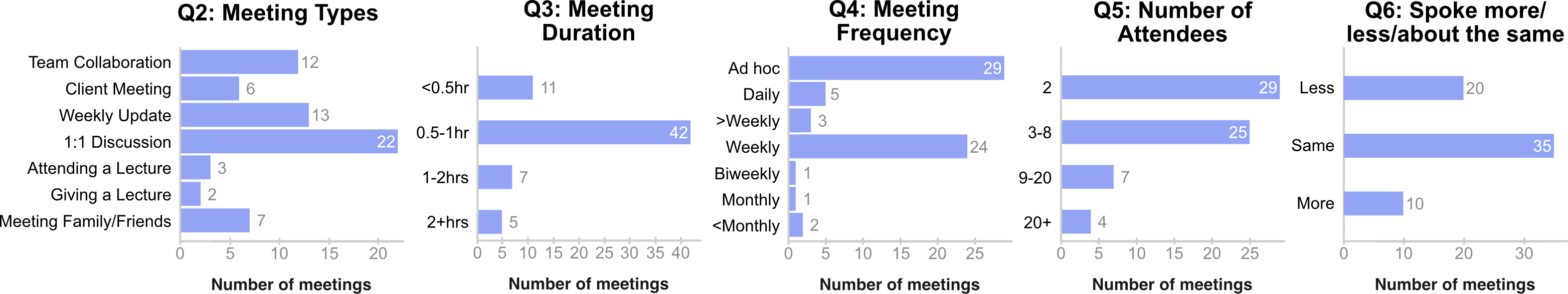} 
    \vspace*{-0.5em}
    \caption{Breakdown of meeting contexts reported in the diaries (65 entries - See also: Supplementary Material).}
    \Description{Bar charts showing the distribution of (Q2) meeting types, (Q3) meeting durations, (Q4) frequency of meetings joined, (Q5) number of attendees, and (Q6) whether participants spoke more, about the same, or less in the meeting.}
    \label{fig:diary_analysis}
    \vspace{-0.8em}
\end{figure}

We collected 65 diary entries, spanning diverse meeting contexts (see \autoref{fig:diary_analysis}). We report on key findings from our analysis of diaries and interviews, organized by research question.

\subsection{Impact of Meaning-Oriented Feedback on Noticeability of Visual Non-Verbal Cues}

Our RQ1 asked: How does real-time feedback on the potential meanings of visual non-verbal cues
help remote meeting attendees \emph{notice} when their behavior aligns or diverges from their
communicative intent? We found that not only did most participants perceive they engaged more frequently and for longer periods of time on their self-view, they recalled a perceived shift in focus from appearance self-checks to  considering how their visual non-verbal cues influence communication. For some, this resulted in a perceived process of more deeply reevaluating how they routinely present themselves to others and the impacts on others' impressions and understanding.

\vspace*{0.3em}
\textbf{Participants noted an increased perceived engagement with \emph{FaceValue}, compared to traditional self-views with no feedback.} 
They felt that using \emph{FaceValue} transformed occasional self-checks into frequent (10/13), and longer (7/13) glances at their self-view. For example, \pid{6} described how \emph{FaceValue} prompted them to monitor themselves during the meeting rather than just during initial setup:
\tinyskip
    \reportquote{``I'm definitely looking at myself more because I think, when I join the meeting [without \emph{FaceValue}], I make sure everything looks OK. And if I say something in the middle of the meeting and I'm not sure where to look, I would just look at myself. But with \emph{FaceValue}, I feel it nudges me to look [at myself] more.''}{\pid{6}}
\tinyskip

\noindent Many participants (8/13) implied that it was \emph{FaceValue}'s subtle nudges that prompted more frequent self-checks. Participants' diaries also indicate that they felt they viewed themselves more often (3.85$\pm$1.01, slight agreement) compared to their regular remote meeting experiences.  Whether this increased engagement with the self-view is purposeful and useful is discussed next. 

\miniskip
\textbf{Participants conveyed the impression that the augmented self-view shifted their attention towards communicative meaning of their visual non-verbal cues.} Participants reported that \emph{FaceValue} led them to pay closer attention to the potential meanings conveyed through facial expressions and head movements (10/13 participants; 46/65 diary entries). They shared:
\tinyskip
    \reportquote{``I think what \emph{FaceValue} added is that, before, when I looked at myself, it was more just for checking how I look. Unless I paid extra attention to my facial expression, I couldn't tell what I was communicating.''}{\pid{11}}
\tinyskip
    \reportquote{``Compared to meetings without \emph{FaceValue}, I now focus on how others will notice my facial expressions and how well I express my emotional feelings to the audience. Thanks to \emph{FaceValue}, I keep my attention on visual non-verbal cues.''}{\pid{9}}
\tinyskip

\noindent Participants felt that meaning-oriented feedback enabled them to notice whether their visual non-verbal cues conveyed meaning in alignment with their communicative intent. For instance, {\pid{6}} and \pid{3} reported using \emph{FaceValue} to help them maintain this self-awareness to ensure they communicated their engagement with the meeting:
\tinyskip
\reportquote{``\emph{FaceValue} allows me to notice when my attention drifts away from looking at other people, which can sometimes translate very poorly to the other side.''}{\pid{6}}
\tinyskip
    \reportquote{``I need to make sure that I'm actively engaged as someone partly responsible for leading the project. So, there were times when I was trying to look at \emph{FaceValue} and get that sense.''}{\pid{3}}
\tinyskip

\noindent Similarly, \pid{1} commented on how \emph{FaceValue} can help with conveyed emotions:

\tinyskip
      \reportquote{``I feel bad if I look grumpy during a meeting. Luckily, I don't think the red box appeared too much. But when it did, I thought: `Okay, maybe I shouldn't look annoyed.' Sometimes it's not that I'm actually angry, but it's how others might perceive it, so I adjusted my expression.''}{\pid{1}}
\tinyskip

\noindent Such an active assessment of how others may possibly misinterpret one's communicative intent suggests that some participants may have begun to counter the Illusion of Transparency: the mistaken belief that one’s internal state is clearly understood by others (\autoref{tab:theory-overview}--T4). Critically, participants expressed that this self-awareness is focused on the communicative meaning implied by their non-verbal facial expressions---even when they are not speaking.

\miniskip
\textbf{Some participants reported engaging in broader reflection on meaning-oriented self-presentation.}
\noindent In-the-moment realizations illustrated above could lead to a more profound reflection on how one's cues might be interpreted more broadly, across meetings, roles, and audiences; and on one's behavioral habits. For instance, {\pid{5}} described how \emph{FaceValue} led them to reconsider the role of visual non-verbal cues in their interactions with others:
\tinyskip
  \reportquote{``I learned a lot about myself. I assumed before \emph{FaceValue} that I wouldn't care about my behavior. I'd probably just sit there blank-faced. But with \emph{FaceValue} I realized the importance of making more of an effort to look engaged.''}{\pid{5}}
\tinyskip
\noindent This suggests that the self-awareness fostered by \emph{FaceValue} helped this participant reflect on how their non-verbal expressions might communicate meaning in everyday interactions. Some participants reported beginning to reevaluate their expressive habits, not just in response to the tool, but in reflection on how they were routinely perceived by others. 

\miniskip
\textbf{Challenge: Algorithmic misapprehensions can hinder insight.} 
Most participants (8/13) described the detection as predictable enough for real-time awareness, and found \emph{FaceValue}'s feedback useful and actionable (8/13; 19/65 diaries). For instance:
\tinyskip
    \reportquote{``One thing I found really helpful is, I was able to see that I look really confused when I was thinking deeply. I think that helped me communicate `this part is a little confusing. Could you explain?' ''}{\pid{7}}
\tinyskip

\noindent However, some noted challenges with \emph{FaceValue}'s interpretation of their cues. \pid{16} reported:

\tinyskip
    \reportquote{``Whenever I'm thinking, it tends to flag me as negative emotion for whatever reason, even if that isn't my intention.''}{\pid{16}}
\tinyskip

\noindent For \pid{12}, the recognition appeared random:
\tinyskip
    \reportquote{``I couldn't figure out what it was trying to detect in terms of positive or negative emotions. It seemed essentially random to me, so I couldn't discern the pattern.''}{\pid{12}}
\tinyskip

\noindent In these instances, participants were unable to reconcile \emph{FaceValue}'s feedback with their cues, and attributed this to difficulties understanding what the algorithm did and why. \pid{16}'s quote suggests that feedback alone was not enough to trigger useful self-awareness, whereas \pid{12}'s points to how unpredictability can prevent meaningful insight altogether.

\miniskip
\textbf{Challenge: Continuous feedback is prone to self-monitoring fatigue.} 
While continuous overlays may enhance awareness for some, they can be tiring for others. Several participants (\pid{3}, \pid{5}, \pid{6}, \pid{16}) reported that \emph{FaceValue} increased fatigue or mental effort due to the extra attention required to interpret its overlays. However, \pid{6} noted that any added load could be positive:
\tinyskip
    \reportquote{``In terms of cognitive load, maybe there's an increase. But again, I think that's a good thing: it means I'm paying more attention to whatever is happening in the meeting.''}{\pid{6}}
\tinyskip
\noindent \pid{7} reported that they deliberately reduced how often they looked at the feedback to avoid becoming overwhelmed. These perspectives suggest variability in how users balance \emph{FaceValue}'s benefits against the mental overhead of continuous self-monitoring. 

\miniskip
\textbf{Summary.} Participants' reports suggest that \emph{FaceValue} usage can support purposeful self-monitoring, encouraging reflection on how visual non-verbal cues might be interpreted by others. For many, this self-awareness was beneficial. However, others, found \emph{FaceValue} unpredictable or hard to interpret, limiting the insight they gained from the tool. Some also reported increased mental fatigue from extra monitoring, raising questions about the tradeoff between providing continuous feedback to improve communication and the cognitive cost of processing such feedback.
\vspace{-0.5em}

\subsection{From Self-Awareness to Action}

RQ2 asked: How do attendees describe attempting to \emph{act} on meaning-oriented feedback to realign their visible cues? Participants reported leveraging \emph{FaceValue}'s feedback to reinforce or course-correct their visual non-verbal cues in real-time, with these adjustments perceived to have a positive impact on others. While \emph{FaceValue}'s suggestive feedback was generally perceived as actionable, participants reported instances where the lack of guidance left them unsure of how to respond.

\textbf{Meaning-oriented adjustment and realignment.} Participants reported adjusting their visual non-verbal cues in response to \emph{FaceValue}'s feedback, aiming to better align cues with their communicative intent. They described taking such actions immediately after noticing a color shift or overlay (6/13; 17/65 diary entries), which alerted them to sudden shifts in their cues.
\tinyskip
    \reportquote{``I was just listening [...] Sometimes I realized I wasn't paying attention, but when I saw a negative indication through \emph{FaceValue}, I corrected my expression so I didn't appear negative.''}{\pid{11}}
\tinyskip
     \reportquote{``If there's a sudden change, especially if there's a red thing that appears, I would immediately glance at myself, and then readjust, right? Correct course. [...] I would interpret the red as, `Oh, there's been a sudden negative shift in my demeanor,' so [I] look at myself, and then immediately readjust.''}{\pid{8}}
\tinyskip
    \reportquote{``I wanted to stay positive, but when \emph{FaceValue} showed yellow, I thought `I don't want to be neutral,' so I changed my expression to be more positive to convey my intent clearly.''}{\pid{9}}
\tinyskip

\noindent These examples illustrate \emph{FaceValue}'s value in identifying unintentional drifts in expression (\pid{11}), misaligned cues (\pid{8}), or signals too subtle to be seen (\pid{9}), allowing immediate adjustments. Participants also reported intentionally amplifying some of their cues to trigger overlays which aligned with their communicative intent. \pid{8} describes such an instance: 
\tinyskip
    \reportquote{``When I nod I could see that I'm nodding with or without [by seeing the lines around my face]. Because of that feedback, I might be [intentionally] nodding more than usual.''}{\pid{8}}
\tinyskip

\textbf{Perceived impacts of adjusting one's visual non-verbal cues on others.} Beyond achieving clearer communication through realignment of their cues, some participants believed they saw positive changes in how others perceived them due to these adjustments. While we could not confirm this perception with other meeting attendees, three participants explicitly connected \emph{FaceValue}-prompted adjustments with improved impressions or communication clarity. For instance:

\tinyskip
    \reportquote{``The interviewer said that I was lovely and brilliant and they were excited to work with me through this process. [...] Due to \emph{FaceValue}, my real-time facial expressions reinforced the enthusiasm that I was vocalizing.''}{\pid{5} (diary entry)}
\tinyskip
    \reportquote{``I was nodding and the \emph{FaceValue} detected that I'm nodding. And I felt like, Yeah, I do want to nod. So I actually nodded more. And my participant, noticed this, as he was definitely smiling at my movement. So that is when I felt [the \emph{FaceValue}] makes a difference.''}{\pid{14}}
\tinyskip

\textbf{Challenge: Awareness may not translate to actionability.} 
While \emph{FaceValue} helped participants with a self-awareness of their expressions---both to help them to notice and correct cues (8/13)---some participants (3/13) explicitly reported uncertainty about how to respond to its prompts:
\tinyskip
    \reportquote{``I didn't pay much attention to the icons. They weren't distracting, but they weren't particularly helpful either.''}{\pid{13}}
\tinyskip
    \reportquote{``When the interface reacts to head movement, is that good or bad? What should I do? Should I move more or less?''}{\pid{16}}
\tinyskip

\noindent These comments are consistent with Chow et al.~\cite{r6_chow2025}, who also found that some participants felt at a loss about what to do in response to the nudges and indications the tool provided.
\miniskip

\noindent\textbf{Summary.} \emph{FaceValue} may help participants steer interactions intentionally through careful use of facial expressions and head movements, potentially reflecting Objective Self-Awareness~\cite{duval1972self} in action. However, some participants did not know whether and how to adjust their cues using feedback alone, raising the question of actionability of suggestive feedback.

\subsection{Continuous Feedback Is Useful... Some of The Time}

Our RQ3 asked: How do meeting characteristics (e.g., stakes, formality) shape how the feedback is perceived in terms of usefulness and distraction? Providing continuous feedback on one's visual non-verbal cues may not always be relevant or appropriate~\cite{chen2010contribution}, and a question that was posed by prior work~\cite{r6_chow2025}. We found that engaging in self-monitoring was particularly important in professional contexts, but less relevant for casual interactions with relatives and friends.

\miniskip
\textbf{Self-Monitoring is deemed important in high-stakes and task-oriented meetings; less so in casual meetings with familiar people.} Participants (7/13) found \emph{FaceValue} especially helpful in high-stakes or task-oriented meetings (e.g., interviews, team projects) where one's demeanor is important. For example, \pid{7} appreciated it during sprints, and \pid{11} in large-scale gatherings:
\tinyskip
    \reportquote{``It was a large-scale meeting with many people. [...] Even if I'm not speaking, if the other person is speaking and I agree, I want to communicate I'm in agreement. So I would monitor myself. I look at [\emph{FaceValue}], and that was helpful.''}{\pid{11}}
\tinyskip
\noindent Similarly, \pid{8} valued the tool in stressful or sensitive discussions:
\tinyskip
    \reportquote{``During stressful meetings, it was especially important for me [to know] how I was coming across to the other person. I wouldn't necessarily have the time to look at my \emph{FaceValue} window as much because I'm so engrossed in the conversation. But then, if there was a sudden change in my expressions, there would be that pop up, the green or the red one. So that was helpful, because it served as a cue for me to readjust. For example, if I was stressed and then I see a red pop up. I was like, `Oh, no, I'm coming across stressed.’ So I would just readjust.''}{\pid{8}}
\tinyskip

\noindent These are suggestive of how the self-awareness provided by \emph{FaceValue} provided clarity to their impression management efforts. \pid{1} found it also helped to support their intentional impression management efforts when encountering unfamiliar people:

\tinyskip
    \reportquote{``I think it can help if you're less familiar with people to make sure that you're appropriately communicating non-verbally.''}{\pid{1}}
\tinyskip

\begin{figure}[!t]
    \centering
    \includegraphics[width=0.7\linewidth]{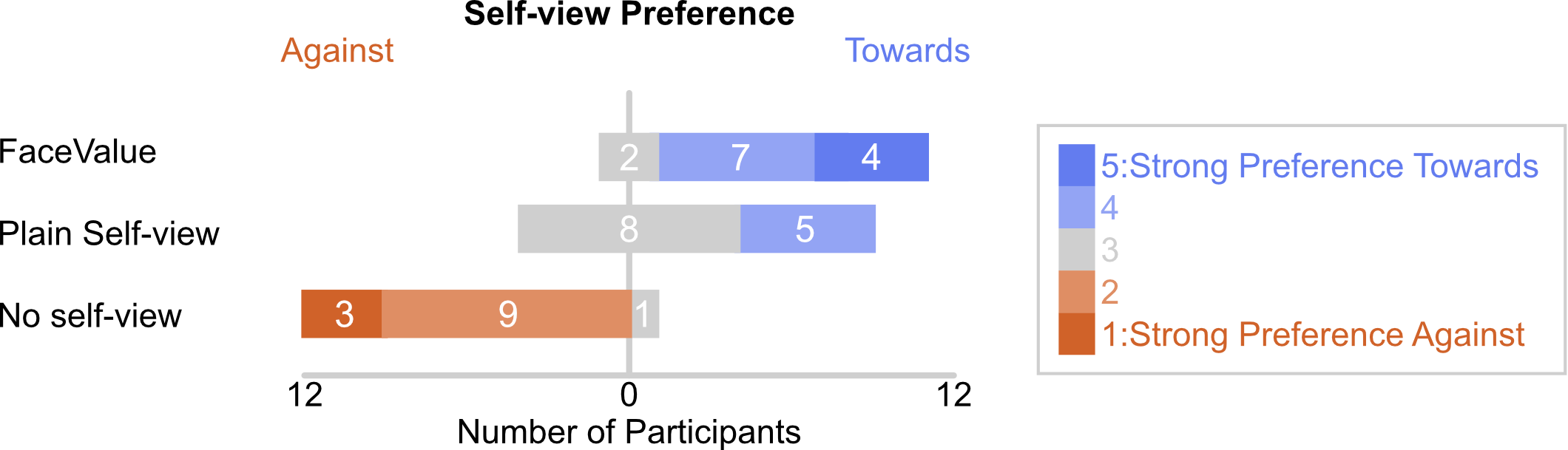}
    \caption{Participants' self-view preferences collected in the post-study questionnaire.}
    \Description{The stacked bar chart shows that 4 participants strongly preferred FaceValue, 7 preferred it, and 2 were neutral. For a plain self-view, 5 participants preferred it, and 8 were neutral. For no self-view, 3 participants strongly preferred against it, 9 were against it, and 1 was neutral.}
    \vspace*{-0.3em}
    \label{fig:pref}
\end{figure}

\noindent Four participants found the feedback less relevant for casual or personal meetings; \pid{9} describes:
\tinyskip
    \reportquote{``If you are in casual meetings, for example, if you're chatting with your friends or family, of course it's important to be genuine and not being rude towards the others but you can be true to yourself.''}{\pid{9}}
\tinyskip
\noindent Overall, even if active self-monitoring is not called for, all but two participants indicated that they would prefer an augmented view over the regular self-view or no self-view (see~\autoref{fig:pref}).

\miniskip
\textbf{Summary.} \emph{FaceValue} was perceived to be the most valuable in meeting contexts where nuanced communication through visual non-verbal cues had potential implications, such as one's making an impression during an interview or contributing to team's clear communication. 

\subsection{Design Reflection}
Our deployment study allowed us to gain insight into the role and impacts of design choices behind \emph{FaceValue}. In the following, we discuss participants' observations in relation to holistic feedback, suggestive (as opposed to prescriptive) prompts, and private overlay design.

\miniskip
\textbf{Support for holistic self-monitoring.} %
By overlaying attendees' faces with comic-style cues derived from multiple signals (\autoref{fig:cues}), our design aimed to encourage holistic, wide-ranged awareness of one's expressions, rather than an analytic one that would isolate instances of particular facial signals~\cite{boehner2005affect}. Ten participants referred to their cues using holistic language rather than analytical language. For instance, \pid{3} said ``I'm looking actively engaged,'' and \pid{10} said ``positive emotion.'' Occasionally, these participants used analytical terms to describe the overall state of their facial expression or head movements. In contrast, \pid{16}, \pid{14}, and \pid{12} mostly described adjusting specific facial parts as their goal, suggesting a more analytical focus. For example, \pid{14} said, ``When I'm nodding and \emph{FaceValue} detects I'm nodding, I feel like I do want to nod, so I actually nod more.''

\miniskip
\textbf{Support for predictable, and trustworthy self-monitoring.} 
Acknowledging FER's limitations, our approach embraces imperfect sensor readings and models to create meaningful user interactions, treating ambiguity as a resource. This was only partially successful: 8/13 participants found detection predictable enough for real-time awareness, despite occasional false positives, while others found the feedback too unpredictable. \pid{16} and \pid{1} noted:

\tinyskip
    \reportquote{``I think whenever I'm confused, \emph{FaceValue} is probably going to show me as angry. [...] Whenever it's not aligning with my actual behavior, I start to not look at it for a while.''}{\pid{16}}
\tinyskip
    \reportquote{``I do think there were maybe a few false positives where, if I tilted my head a little bit, it would think I was shaking my head, but it would usually disappear pretty quickly. Because all these reactions are transient, I think the false positives don't really matter that much. And I think you can naturally build that in and just wait a little longer [for effects to disappear].''}{\pid{1}}
\tinyskip

\noindent Although \emph{FaceValue} was positively perceived with respect to trustworthiness (13/13) overall, there are still isolated instances where feedback was too unclear to be actionable. For some, it was easy to ignore, especially since this feedback was private. For two participants, that could have contributed to undermine the perceived value of such system: 2 participants (\pid{16} and \pid{12}) rated the plain self-view higher (slight preference towards), than \emph{FaceValue} (rated neutral) for remote meetings; with \pid{12} rating no self-view also neutrally, whereas \pid{16} reported a strong preference against no self-view. 
These two participants explicitly described \emph{FaceValue} as distracting (\pid{5}, \pid{12}) primarily in relation to specific low-stakes or informal meetings where they felt it prompted more self-monitoring than they considered necessary. Otherwise, accuracy and actionability issues did not appear to undermine the overall usefulness of \emph{FaceValue}. These accounts are consistent with our intention to use ambiguity as a design resource: for many participants, the overlays' ambiguity seemed to make FER inaccuracies tolerable rather than fatal to the tool's perceived usefulness.

\textbf{Preserving user privacy with private overlay} 
We used private overlays for the visual effects, so only could see feedback on their visual non-verbal cues. As a result, privacy concerns around privacy did not emerge as a major barrier. Ten  participants reported no issues under those conditions. \pid{6} and \pid{1} summarized:

\tinyskip
    \reportquote{``If the [\emph{FaceValue} view] that I'm seeing is also going to be sent to the recipient, I might have second doubts about that. I feel like, ‘Oh, it's a gauge that I can choose to ignore, or I can choose to use that information to make sure that I'm still in control.' ''}{\pid{6}}
\tinyskip
    \reportquote{``I was initially a bit concerned about privacy, but I think the local setup alleviated a lot of that.''}{\pid{1}}
\tinyskip

\noindent Some (3/13) still acknowledged hypothetical risks if overlays were inadvertently shared or if organizers could access \emph{FaceValue} data, underscoring the importance of user control over visibility.

\section{Discussion}
\label{sect:discussion}

Our findings highlight \emph{FaceValue}'s potential to evoke meaning-oriented self-awareness via real-time feedback on facial expressions and head movements. We discuss these results in relation to existing theories, outline implications for future systems, and review our study's scope and limitations.

\subsection{Suggesting Nuances to Objective Self-Awareness: From Sensing to Adjusting}

Our participants reported that the external cues (overlayed feedback) made monitoring feel more frequent, and shifted their attention to what was being monitored. They described reflecting more on how their expressions might be interpreted by others.
Interpreting this through the Objective Self-Awareness (OSA) theory, we view \emph{FaceValue} as an example of a computational tool that could support a shift in self-monitoring from cosmetic checking to meaning-oriented inspection (\textbf{DG1}), via continuous, real-time overlays (\textbf{DG3}). In contrast, prior OSA work has emphasized internal self-evaluation upon noticing oneself as a social object. 

Further, real-time feedback was experienced as a bridge between internal awareness and expressive control, which participants described prompted them to ``course-correct'' their visual cues in the moment. From the perspective of OSA, this suggests that computation tools could serve as a behavioral mediator, helping users apply OSA in situ.

This is consistent with findings from Chow et al.~\cite{r6_chow2025}, where a glanceable sidebar tallying detected smiles, nods, gaze, and postures increased self-awareness. According to our participants' reports, when feedback is more aligned with the rich, interpretive nature of visual non-verbal cues, attendees felt less need to fixate on a single trigger. These reports complement the insights of Chow et al., where %
 participants typically responded with narrower actions like ``don't smile'' or ``nod more often.'' Chow et al. were motivated by reducing `non-verbal overload' and Zoom fatigue, which resulted in favoring an analytic, count-based approach. In contrast, our participants used vocabulary describing their goals and intentions at a higher level. For instance, \pid{3} reported wanting to appear more ``engaged,'' and \pid{10} to express more ``positive emotion.'' The design approach we took in \emph{FaceValue} invited participants to consider holistic interpretations of the communicative, meaning-oriented aspects of the non-verbal behavior. Participants reasoned in terms of ``meaning'' rather than  easily-countable behaviors that sometimes serve as proxies for these meanings.

Further, privacy-preserving design choices, such as on-device processing and private overlays, were highly valued. Guaranteeing users full control over who sees feedback and when can reduce the emotional stakes, as users can choose to engage or ignore the feedback.

\vspace{0.5em}
\reportimplication{
\textbf{Design Implication 1:} Subtle visual feedback can increase the perceived communicative usefulness of self-monitoring. Systems should support both detection (awareness) and enactment (adjustment) of meaningful cues through private feedback that preserves users agency and emotional safety.
}

\subsection{Disrupting the Illusion of Transparency Through Reflective Awareness}

Participants' comments indicated that \emph{FaceValue} disrupted a tacit assumption that their communicative intent was obvious to others. This can be interpreted as supporting and extending prior research on the Illusion of Transparency, which argues that people overestimate how well their internal states are perceived externally. Rather than directly reporting overconfidence, our participants retroactively realized how their expressions might have been misinterpreted, implying that the illusion may stem not from confidence but from lack of attention. This addresses RQ1.

We thus posit that raising awareness of interpretation possibilities (\textbf{DG1}) may be an important precursor for breaking the illusion. \emph{FaceValue} did this without judgment or exposure, suggesting a direction for feedback tools that improve communication clarity without increasing pressure.

\vspace*{0.5em}
\reportimplication{
\textbf{Design implication 2:} Feedback tools can potentially help break the illusion of transparency, by prompting reflection on how visual non-verbal cues might be interpreted by others.
}
\subsection{Interpreting Feedback: How Individual Users Made Sense of \emph{FaceValue}}
\label{sec:interpretation}

Participants varied in how they interpreted and responded to feedback from \emph{FaceValue}. Most participants found \emph{FaceValue} predictable enough to form their own interpretations (\textbf{DG2}), although others struggled with interpreting its feedback. These responses echo Gaver's notion of \emph{ambiguity as a resource for design}~\cite{gaver2003ambiguity}, in which partial information can be valuable if it invites interpretation. This design ambiguity seemed to let participants engage more deeply with \emph{FaceValue}'s overlays, enabling participants to form more holistic, wider-scoped meanings %
such as ``looking engaged,'' ``positive''. %
As expected, while our results show that some participants saw ambiguity as a prompt for reflection, others perceived it as a failure of clarity. Therefore, designers would need to carefully consider user background and cognitive load, perhaps offering modes that adjust granularity or explanation level, concrete guidance, or pre-meeting training, to accommodate different user needs. Pre-meeting training could particularly focus on establishing a common understanding of what constitutes effective visual non-verbal cues in communication, thereby improving the actionability of the feedback provided by \emph{FaceValue}.

Overall, participants expressed a strong preference for \emph{FaceValue} over both a plain self-view and no self-view. This suggests that subtle, suggestive feedback, when private, can be more desirable than both absence and crude monitoring. However, our findings also suggest that for serving a broader population, systems should consider users' interpretive bandwidth. %
Future systems may benefit from adaptive modes that align with user needs, preferences, and cognitive load.%

\vspace*{0.5em}
\reportimplication{
\textbf{Design implication 3:}  Effective meaning-oriented self-monitoring systems should dynamically adapt to users' momentary interpretive bandwidth.}%

\subsection{
Context-Dependent Utility}

The utility of \emph{FaceValue} varies by context. Participants emphasized that \emph{FaceValue} was most helpful in formal or high-stakes meetings. This context-dependent perception of utility addresses RQ3 directly. This finding aligns with the Media Richness Theory~\cite{daft1986organizational}, which suggests that ambiguous and complex tasks benefit from richer communication channels. Taken together, our results illustrate this theory by exemplifying how real-time refinements of visual non-verbal cue may help offset communication richness from the cueing attendee's perspective. 

Participants also found \emph{FaceValue} helpful in relationally ambiguous contexts (e.g., with unfamiliar participants), which aligns with the Social Information Processing (SIP) theory~\cite{walther1992interpersonal}. They reported little use in casual conversations, suggesting that relational stakes modulate perceived utility.

Taken together, these findings suggest that \emph{FaceValue} is most valuable when (1) the stakes are high, (2) relational dynamics are uncertain, or (3) tasks require clear, nuanced communication. In casual or low-stakes settings, however, \emph{FaceValue} risks feeling intrusive or distracting (``Clippy-like'').

The Task-Technology Fit (TTF) theory~\cite{goodhue1995task} helps explain  participants' impressions. In low-stakes situations, task demands are also low, so the perceivable benefit of launching and attending to \emph{FaceValue} is likely small. In this case, the activation cost may seem high and users may choose to forgo using \emph{FaceValue}. In contrast, in high-stakes or ambiguous situations, the benefits offered by \emph{FaceValue} may be deemed to outweigh the activation cost, making its use appear worthwhile.

Guided by the TTF, we can also avoid ``Clippy-like'' intrusion. Our results suggest that participants encountered task characteristics in real-world meeting scenarios that match the technology's characteristics, yielding task-technology fit. Under high-fit conditions, the theory predicts (1) greater utilization and (2) improved performance impacts (conditional on use), which are effects we desire to observe. Accordingly, \emph{FaceValue} should be surfaced and easy to invoke when task characteristics match the technology's capabilities, and otherwise should fade to avoid adding unnecessary complexity.

Designers should therefore ensure that feedback is context-sensitive and adaptive, appearing only when it supports the user’s goals and enhances, rather than disrupts, the flow of conversation.

\vspace*{0.5em}
\reportimplication{
\textbf{Design implication 4:} Effective meaning-oriented self-monitoring systems should strive to encourage more authentic visual non-verbal expressions while dynamically adapting to users' situational context.
}

\subsection{Ethical Concerns}
Several ethical considerations arise when providing feedback on facial expressions in remote meetings. In this section, we outline key concerns and discuss how the design choices in FaceValue aim to address, complicate, or limit their impact.

\textbf{Cultural normativity of FER expression labels.} Facial-expression recognition (FER) systems often rely on labels derived from specific cultural contexts, which can impose a normative interpretation of visual cues. This risks flattening culturally diverse forms of expressivity into a single reading. FaceValue mitigates this concern by using intentionally color-based overlays rather than fixed emotion labels, allowing users to interpret the feedback in ways that fit their own communicative norms. However, ambiguity does not eliminate underlying cultural assumptions in the training data. Future work should explore culturally-situated calibration or user-tailored~interpretations.

\textbf{Neurodivergent and atypical expressivity.} Many neurodivergent individuals, including those on the autism spectrum, experience difficulties in producing or interpreting facial expressions~\cite{trevisan2018facial}. Systems that imply a ``typical'' expressive baseline may inadvertently reinforce norms that do not fit all users. FaceValue's reflective cues may help illuminate how their expressions could be perceived by others in remote settings. At the same time, such feedback may increase pressure to conform to neurotypical expressions, potentially contributing to emotional labor or masking. Further work is needed to understand when, and for whom, such feedback is supportive rather than burdensome.

\textbf{Authenticity, performativity, dependency, and emotional labor.}
Highlighting moments of potential misalignment can support users who wish to communicate with greater clarity in high-stakes settings. Yet feedback on one's appearance can also encourage performative self-monitoring, contributing to polished or ``camera-ready'' behavior that may feel less authentic. While FaceValue is designed as a reflective aid rather than a corrective tool, it may still prompt increased self-awareness or emotional labor~\cite{mann1997emotional}, particularly during long or demanding meetings~\cite{bailenson2021nonverbal}. As such, users may develop dependencies, in which they lose their innate self-monitoring skills and refuse to attend meetings without \emph{FaceValue}. Systems that seek to enhance clarity must therefore balance this goal with the need to support authentic and sustainable self-presentation.

Together, these concerns position FaceValue as a voluntary, reflective aid rather than a source of ground truth about one's appearance. Any deployment should make the system's ambiguity, limitations, and optional nature explicit, ensuring users can decide whether such feedback~is~helpful, and can easily opt in or out depending on the interaction context. Similarly, deployments could present the system (or its feedback) less frequently, framing it as a training tool rather than an always-available mirror.

\subsection{Limitations and Future Work}
\label{sect:futurework}

Our current study focused on $n=13$ participants, most of whom were students and researchers comfortable with self-view. Future work should include participants with more diverse demographics, professional roles, and comfort levels with video, enabling the study of non-normative behaviours (e.g., camera-off attendees~\cite{balogova2022you}) and testing whether meaning-oriented feedback supports or disrupts communication in different contexts.

While we obtained rich insights from participants addressing our research questions, we focused on internal instruments to collect subjective qualitative data obtained through one side of the conversation. As such, we acknowledge the limitations of not providing objective evidence of changes or group-level effects. 
Our participants may over- or understate effects, or misread others' reactions, and their accounts should be read as subjective, situated interpretations rather than ground truth about behavior. The research had to rely only on theory-based speculation and participants' direct quotes to gauge how many benefits \emph{FaceValue} provided in terms of improving understanding of cues or group dynamics.

To address this limitation, future work could capture richer behavioral data and examine long-term use. Researchers could explore privacy-preserving hybrid deployments that collect richer interaction data under controlled conditions, allowing for detailed analysis of how participants actually engage with feedback in real time. Longitudinal studies across varied meeting types could also investigate how attitudes and behaviors change over time and whether sustained exposure leads to lasting improvements in self-awareness and cue adjustment. Similarly, future work could go beyond self-reports and examine group-level effects. Evaluations could triangulate self-reported findings with external measures (e.g., other meeting participants' direct feedback, conversation outcomes, or third-party ratings of communication effectiveness) to provide stronger evidence of impact. This would extend the perspective beyond the focal user and reveal how feedback systems influence group dynamics, shared understanding, and meeting~outcomes.

Finally, future work could explicitly investigate individual differences in trait-level self-monitoring and how they interact with the behaviors that \emph{FaceValue} supports. For example, people high in trait self-monitoring may experience \emph{FaceValue} as a helpful resource for fine-tuning their self-presentation, whereas people low in trait self-monitoring may find the same feedback distracting or unnecessary. Although our study did not measure this trait directly, our findings also suggest that some participants cared more about ongoing self-monitoring than others. These differences in trait-level self-monitoring may help explain why some participants disliked using \emph{FaceValue}. Measuring trait self-monitoring alongside behavioral and experiential outcomes would allow us to account for this confound in preference testing. Such measurement would also enable us to estimate the effect size of trait self-monitoring on behavior change, relative to baseline measures. Similarly, future work could explore how the system might be tuned or calibrated for users from different cultural backgrounds or for neurodivergent users.

\section{Conclusion}

This study introduced \emph{FaceValue}, a lightweight self-view augmentation designed to support meaning-oriented self-monitoring in remote meetings. Building on theories of Objective Self-Awareness, Illusion of Transparency, and Social Information Processing, we explored how real-time, suggestive feedback on facial expressions might enhance attendees' perceived awareness and adjustment of visual non-verbal cues.

Our in-the-wild deployment with thirteen knowledge workers revealed that participants not only perceived themselves to self-monitor more frequently, but also perceived they had shifted their focus from cosmetic appearance to the communicative meanings of their cues. This heightened awareness, in many cases, may have translated into intentional real-time adjustments, improving how participants felt about how they might be perceived by other attendees. Importantly, interpretation of the feedback varied by user and context (which points to selective, context-dependent deployment, rather than as an always-on solution). While high-stakes and goal-oriented meetings amplified \emph{FaceValue}'s utility, casual or familiar conversations often rendered it less relevant. Participants also emphasized the importance of design features, such as privacy preservation in developing remote meeting support tools. We present four design implications based on our study's results that can inform future designers of remote meeting systems.

\begin{acks}
This research was supported by Meta Reality Labs, the
National Science and Engineering Research Council (RGPIN-2018-05072), and the Faculty of Information and the Department of Computer Science
(University of Toronto).
We would also like
to thank the members of the Dynamic Graphics Project for their
valuable advice and assistance, as well as the participants of the
user study for their contributions. We also thank Prof.~Jens Emil Sloth Gr{\o}nb{\ae}k for helpful discussions at an early stage of this project.

 We acknowledge the use of ChatGPT by OpenAI on proofreading the text, and refactoring/optimizing the code for developing FaceValue.

\end{acks}

\bibliographystyle{ACM-Reference-Format}
\bibliography{OutsideZotero}

\end{document}